\documentclass[newversion]{aa}
\usepackage{graphicx}

\begin{document}


   \title{Dependence of the He/H$_8$ emission ratio on brightness, \\
temperature, and structuring of prominences}

   \author{G. Stellmacher\inst{1}
          \and E. Wiehr\inst{2}}

   \offprints{E. Wiehr}

   \mail{ewiehr@astrophysik.uni-goettingen.de}

   \institute{Institute d'Astrophysique (IAP), 98 bis Blvd. d'Arago, 
              75014 Paris, France
              \and
              Institut f\" ur Astrophysik der Universit\"at,
              Friedrich-Hund-Platz 1, 37077 G\"ottingen, Germany}

   \date{Received July 14, 1993; accepted Dez. 11, 1993}

\abstract
{}
{We investigate the dependence of the He/H$_8$ emission ratio on 
kinetic temperature and total Balmer brightness.}
{The line pair He\,{\sc i}\,3888\,\AA{} and H$_8$\,3889 has been 
observed simultaneously with the Ca\,{\sc ii}\,8498 line in a 
number of quiescent prominences.} 
{The He/H$_8$ emission ratio R is found to cover defined parts of a 
general anti-relation with the total H$_8$ emission, depending on 
the kinetic temperature, $T_{kin}$, of the individual prominence: 
High H$_8$ brightness is related to small R and $T_{kin}$ values, 
and preferably occurs in prominences with a less significant 
fine-structure.} 
{}

\maketitle

%
%

\section{Introduction}

In a previous paper (Stellmacher, Wiehr \& Grupe 1992) it was shown 
that the emission ratio of the lines He\,{\sc i}\,3888\,\AA{} and 
H$_8$\,3889\,\AA{} within one prominence is strongly related to 
the value of the non-thermal line broadening parameter. This 
was explained by the higher (as compared to H$_8$) sensitivity 
of the EUV He-absorption coefficient to the non-thermal line 
broadening. The general decrease of the emission ratio with 
increasing H$_8$ emission, observed for a great number of 
prominences by Moroshenko (1974) and by Illing et al. (1975) 
was not discussed, since the range of observed H$_8$ emissions 
is too small within the single prominence observed in 1992.

Radiative transfer calculations by Moroshenko (1974) show the 
important role of the external Lyman continuum radiation on the 
actual He/H emission ratio. This ionizing Lyman continuum 
radiation directly penetrates in the prominence fine-structure 
('threads') thus depending on their ’structure coefficient’. 
In her calculations, Moroshenko (1974) assumes a constant 
electron temperature, $T_e = 7500$\,K which, however, does 
not hold for faint outer prominence edges, where higher 
temperatures are measured (Hirayama 1971). A correct 
interpretation of the He/H$_8$ emission ratio eventually 
allows the determination of the solar He abundance.

The aim of the present paper is to extend our former 
observations of the He/H$_8$ emission ratio over a much 
wider range of H$_8$ emissions from different prominences, 
and to relate them to the mean kinetic temperature of each 
prominence.

\section{Observations data reduction}
    
We analyzed 5 prominences observed on July 5 and 8, 1992, 
with the Gregory Coud\'e Telescope at Tenerife. The line pair 
He\,{\sc i}\,3888\,\AA{} and H$_8$\,3889\,\AA{} was taken 
in the 5\,th grating order simultaneously with the 
Ca\,{\sc ii}\,8498\,\AA{} line in the 6\,th order using a
spectrum cutter. The spectrograph slit of correspondingly 
1.5\,arcsec width was oriented perpendicular to Earths horizon, 
i.e. in the direction of the (wavelength dependent) atmospheric 
refraction. 

Spectra were taken with a 1024\,x\,1024 pixel CCD camera operated 
at a 2-pixel binning, corresponding to 0.318\,arcsec\,x\,5.9\,m\AA{} 
in the violet, resp. x\,12.9\,m\AA{} in the infrared. In spite of a
12\,sec integration time, the constant seeing conditions allowed 
a spatial resolution of about 2\,arcsec, which is outstanding
for highly resolved prominence emission spectra. Accordingly, 
individual emission streaks were analyzed with a 2\,arcsec spatial 
mean.

Line widths and integrated line emissions ('line radiance') were 
obtained using Gaussian fits. The absolute calibration of the 
violet line pair was performed by comparison with disc center 
intensities at $\lambda = 3888.9$\,\AA{}, which are assumed to amount 
to 27\% of the true continuum intensity of $I_{abs}=4.5\cdot10^6$ 
erg/(s\,cm$^2$\,ster\,\AA{} at 3999.0\,\AA{} (Labs \& Neckel 1970). 
Kinetic temperatures, $T_{kin}$, were determined by comparison of  
the H, He, Ca lines as done by Bendlin, Stellmacher and Wiehr (1988).

%

   \begin{figure*}     
   \hspace{5mm}\includegraphics[width=15.8cm]{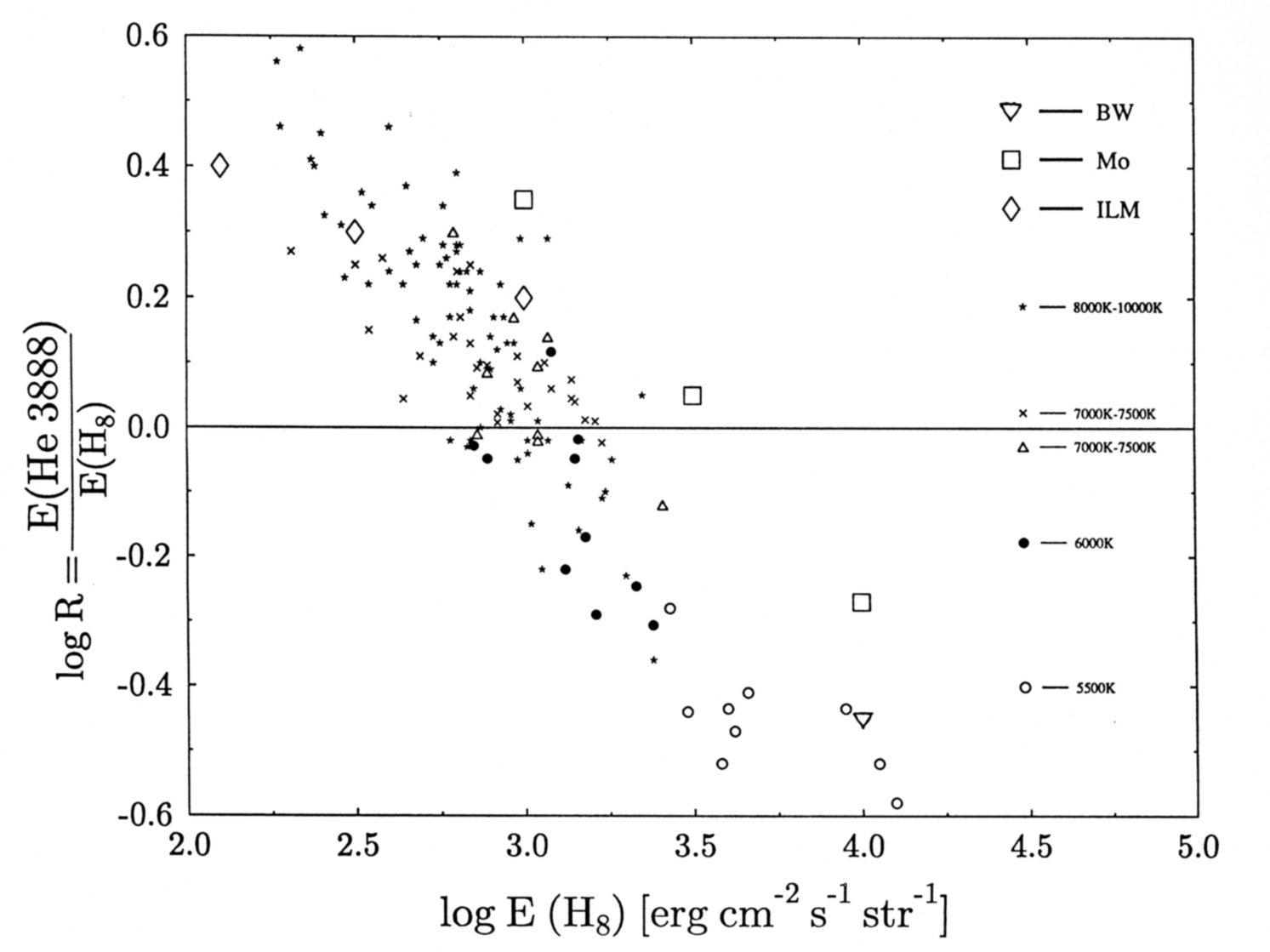}
   \caption{Emission ratio R of the violet line pair 
He\,{\sc i}\,3889 and H$_8$\,3888 versus the total emission 
of the H$_8$\,3888 line. The different prominences are marked 
by different symbols. The corresponding mean $T_{kin}$ value 
of each prominence is given at the right side together with 
the corresponding symbol. Large symbols refer to observed 
means by Moroshenko (Mo), Illing et al. (ILM), Balthasar \& 
Wiehr (BW).}
   \label{Fig1}
    \end{figure*}

\section{Results}   
   
The finally obtained radiance ratio of the He and H$_8$ emission, 
R, on the H$_8$ line radiance is shown in Fig. 1 for the 
different prominences (indicated by different symbols) together 
with their corresponding mean kinetic temperature (right side 
of Fig.\,1). Prominences showing less significant structuring 
in the slit-yaw images are indicated in Figure\,1 by roundish 
symbols. They show highest H$_8$ line radiance, smallest radiance 
ratio R, and smallest mean kinetic temperature $T_{kin}$. In 
turn, the highly structured prominences (non-roundish symbols
in Fig.\,1) show lowest H$_8$ radiance, largest R and largest 
$T_{kin}$ values. The different prominences cover defined parts 
of a general relation, corresponding to their individual (mean) 
H$_8$ radiance and kinetic temperature. 

Comparison with existing data shows that the data by Moroshenko
(1974) are systematically above ours; those by Illing et al. 
(1975) show a different slope (cf., Fig.\,1). We checked that 
these differences are {\it not} due to the higher spatial 
resolution achieved in our spectra: If we simulate the
20\,arcsec spatial resolution achieved by Illing et al. 
(1975), we neither obtain their different gradient in 
Figure\,1 nor the displacement of Moroshenkos data. In 
addition, recent data by Balthasar \& Wiehr (1993) from 
a very bright prominence agree with the lower right part 
of our relation Fig.\,1.

\section{Discussion}

In order to correctly interpret the He/H$_8$ line radiance 
ratio, one has to consider the actual kinetic temperature 
$T_{kin}(\approx T_e)$, the non-thermal broadening and the 
’structure coefficient’ (Moroshenko 1974). The latter parameter 
strongly controls the penetration of the He ionizing EUV radiation 
(Zharkova \& Borchtchevskii 1993). We find that highest H$_8$ 
radiance, smallest emission ratio R, and smallest mean kinetic 
temperature $T_{kin}$ preferably occur in less structured 
prominences. In turn, highly structured prominences show 
lowest H$_8$ raadiance, largest R, and largest $T_{kin}$.
This behavior qualitatively agrees with calculations by 
Zharkova (1989) which require a slightly higher structure
density for brighter prominences.

Without anticipating detailed model calculations, we thus 
argue that the lower temperatures of cool prominences rather 
require a denser packing of threads which, in turn, reduces
the penetration of the He-exciting EUV radiation.

\begin{acknowledgements}
The Gregory Coudé Telescope is operated by the Universit\"ats Sternwarte,
G\"ottingen, at the Spanish Observatorio del Teide of the Instituto de 
Astrof\'isica de Canarias.
\end{acknowledgements}
%
%


\begin{thebibliography}{}

\bibitem[]{}Mattig, W., Schr\"oter, E.\,H.: 1964, Astrophys. J. 140, 804 

\bibitem[]{}Balthasar H., Wiehr E.: 1993 A \& A (submitted)

\bibitem[]{}Bendlin C., Stellmacher G., Wiehr E.: 1988, A \& A 197, 274 

\bibitem[]{}Hirayama, T.: 1971, Sol.Phys. 17, 50

\bibitem[]{}Illing,\,R.\,M.\,E., Landman,\,D.\,A., Mickey,\,D.\,L.: 1975, Sol.Phys. 45, 339 

\bibitem[]{}Labs D., Neckel H.: 1970, Solar Phys 15, 79 

\bibitem[]{}Moroshenko N.\,N.: 1974, Solar Phys 39, 349 

\bibitem[]{}Stellmacher. G., Wiehr, E., Grupe, D.: 1980, A \& A 265, 781 

\bibitem[]{}Zharkova V.\,V.: 1989, in Ruzdjak \& Tandberg-Hanssen (eds.) 
'Dynamics of Quiescent Prominences', IAU-coll. No 117, Hvar Obs. Bull 13, p.331

\bibitem[]{}Zharkova V.\,V., Borcchtchevskii A.\,V.: 1993, Abstract Booklet 
IAU-coll. No. 114, 'Solar Coronal Structures', Tatranska Lomnica

\end{thebibliography}
\end{document}